%% file: template.tex
\documentclass{vgtc}                          % final (conference style)
\ifpdf%                                % if we use pdflatex
  \pdfoutput=1\relax                   % create PDFs from pdfLaTeX
  \usepackage{graphicx}                % allow us to embed graphics files
  \DeclareGraphicsExtensions{.pdf,.png,.jpg,.jpeg} % for pdflatex we expect .pdf, .png, or .jpg files
\else%                                 % else we use pure latex
  \usepackage{graphicx}                % allow us to embed graphics files
  \DeclareGraphicsExtensions{.eps}     % for pure latex we expect eps files
\fi%

\usepackage{microtype}                 % use micro-typography (slightly more compact, better to read)
\PassOptionsToPackage{warn}{textcomp}  % to address font issues with \textrightarrow
\usepackage{textcomp}                  % use better special symbols
\usepackage{mathptmx}                  % use matching math font
\usepackage{times}                     % we use Times as the main font
\usepackage{cite}                      % needed to automatically sort the references
\usepackage{tabu}                      % only used for the table example
\usepackage{booktabs}                  % only used for the table example
\usepackage{enumitem}
\usepackage{footmisc}
\usepackage{graphicx}
\usepackage[table,xcdraw]{xcolor}
\onlineid{0}

\vgtccategory{Research}

\vgtcinsertpkg

\makeatletter
\let\@fnsymbol\@arabic
\makeatother
\title{Inferring Private Personal Attributes of Virtual Reality Users\\ from Head and Hand Motion Data}

\author{
Vivek Nair\thanks{e-mail: vcn@berkeley.edu}\\ %
\scriptsize UC Berkeley\\ %
\and Christian Rack\thanks{e-mail: christian.rack@uni-wuerzburg.de}\\ %
\scriptsize University of Würzburg %
\and Wenbo Guo\thanks{e-mail: henrygwb@berkeley.edu}\\ %
\scriptsize UC Berkeley %
\and Rui Wang\thanks{e-mail: ruiwang813@berkeley.edu}\\ %
\scriptsize UC Berkeley %
\and Shuixian Li\thanks{e-mail: shuixian.li@berkeley.edu}\\ %
\scriptsize UC Berkeley %
\and Brandon Huang\thanks{e-mail: zhaobin@berkeley.edu}\\ %
\scriptsize UC Berkeley %
\and Atticus Cull\thanks{e-mail: atticuscull@berkeley.edu}\\ %
\scriptsize UC Berkeley %
\and James F. O'Brien\thanks{e-mail: job@berkeley.edu}\\ %
\scriptsize UC Berkeley %
\and Marc Latoschik\thanks{e-mail: marc.latoschik@uni-wuerzburg.de}\\ %
\scriptsize University of Würzburg %
\and Louis Rosenberg\thanks{e-mail: louis@unanimous.ai}\\ %
\scriptsize Unanimous AI %
\and Dawn Song\thanks{e-mail: dawnsong@berkeley.edu\vspace{-6em}}\\ %
\scriptsize UC Berkeley %
}

\abstract{
Motion tracking ``telemetry'' data lies at the core of nearly all modern virtual reality (VR) and metaverse experiences. While generally presumed innocuous, recent studies have demonstrated that motion data actually has the potential to uniquely identify VR users.
In this study, we go a step further, showing that a variety of private user information can be inferred just by analyzing motion data recorded from VR devices. We conducted a large-scale survey of VR users (N=1,006) with dozens of questions ranging from background and demographics to behavioral patterns and health information.
We then obtained VR motion samples of each user playing the game ``Beat Saber,'' and attempted to infer their survey responses using just their head and hand motion patterns.
Using simple machine learning models, over $40$ personal attributes could be accurately and consistently inferred from VR motion data alone. Despite this significant observed leakage, there remains limited awareness of the privacy implications of VR motion data, highlighting the pressing need for privacy-preserving mechanisms in multi-user VR applications.
} % end of abstract

\CCScatlist{
  \CCScatTwelve{Security and privacy}{}{}{};
  \CCScatTwelve{Human-centered computing}{Human computer interaction (HCI)}{Interaction paradigms}{Virtual reality};
  \CCScatTwelve{Computing methodologies}{Machine learning}{}{}
}

\begin{document}


\firstsection{Introduction}

\maketitle

\input{100-Introduction}
\input{200-Related-Work}
\input{300-Dataset}

\eject

\input{400-Method}

\eject

\onecolumn
\input{510-TAB-Results}
\twocolumn

\input{500-Evaluation}

\input{600-Discussion}
\input{700-Conclusion}

% \onecolumn

\acknowledgments{We appreciate the help of Beni Issler, Charles Dove, Ines Bouissou, Jason Sun, Julien Piet, Allen Yang, and Xiaoyuan Liu.
This work was supported by the National Science Foundation, the National Physical Science Consortium, the Fannie and John Hertz Foundation, and the Center for Responsible, Decentralized Intelligence.
}

\bibliographystyle{abbrv-doi}
\bibliography{template}

\appendix

\section{Hyperparameters}

\label{app:params}

\begin{itemize}[leftmargin=*]
    \itemsep 0em
    \item Input Shape: $(1024\times21)$
    \item Embedding Size: 24
    \item Hidden Size: 128
    \item Number of Layers: 2
    \item Output Size: 1
    \item Learning Rate: 0.00002
    \item Epochs: 100
    \item Batch Size: 32
\end{itemize}

\clearpage

\newcommand{\true}[1]{\textcolor{orange}{#1}}
\newcommand{\false}[1]{\textcolor{blue}{#1}}

\twocolumn[
\section{Survey Questions}
\label{app:survey}
]

\input{900-Survey}

\twocolumn[
\section{Response Distributions \hfill KEY: \true{CLASS A} -- \false{CLASS B}}
\label{app:dist}
]

\input{910-Dist}

\onecolumn
\clearpage

\section{Response Correlations}
\label{app:corr}

\begin{figure}[!h]
\includegraphics[width=\linewidth]{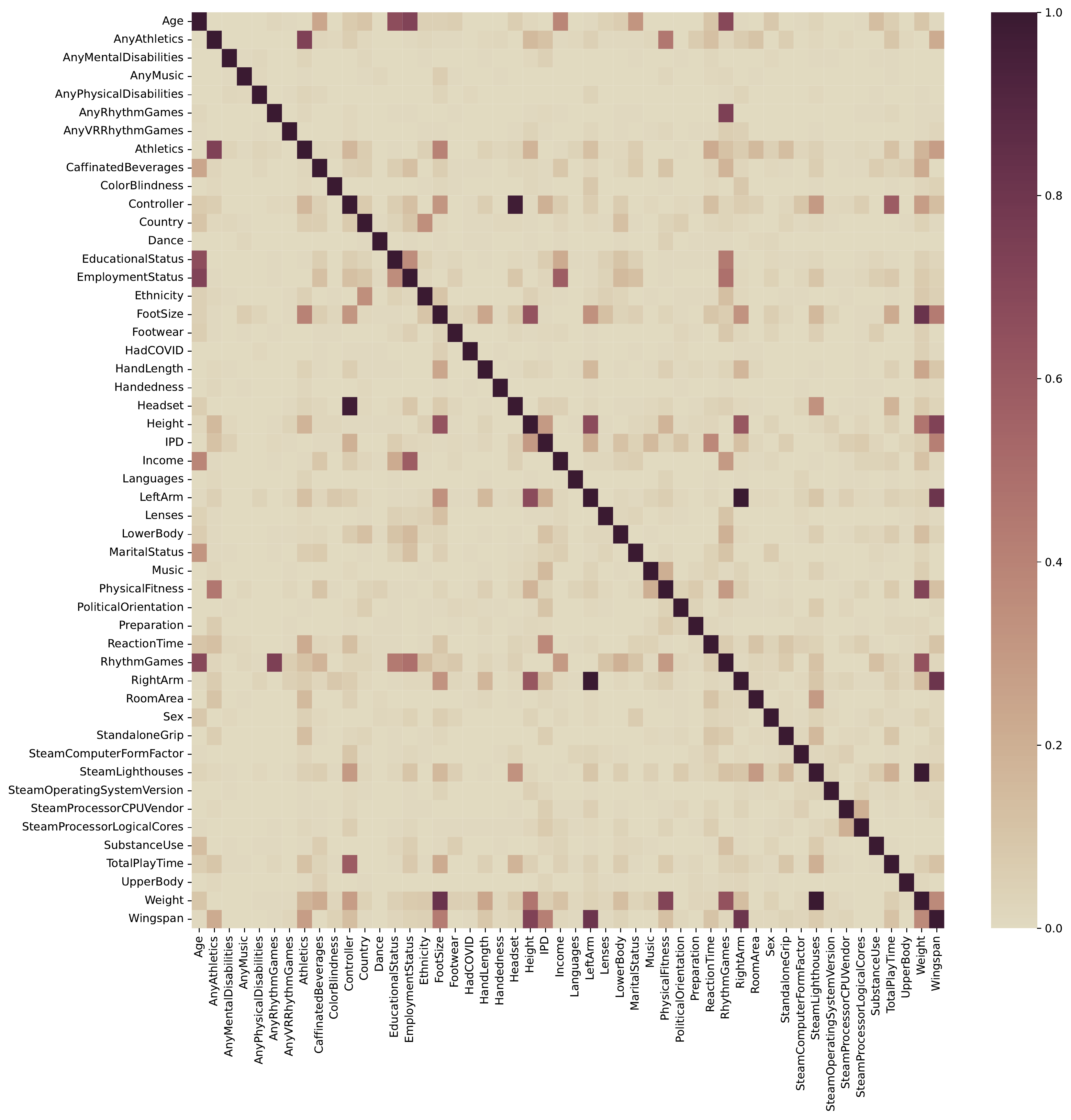}
\centering
\caption{Correlation coefficient ($R^2$) between all pairs of attributes.}
\label{fig:correlations}
\end{figure}

\end{document}

%% file: 100-Introduction.tex
With the recent emergence of affordable standalone virtual reality (VR) devices like the Meta Quest 2, VR technology has begun to reach mass-market adoption for the first time, with nearly 10 million VR systems sold just in 2022 \cite{vr_sales}. While the major proponents of VR envision the ultimate use of these devices to create an immersive virtual ``metaverse'' where users meet to work, learn, and socialize, contemporary adoption of VR has been driven primarily by gaming. As of early 2023, VR games, including ``Beat Saber,'' constitute 91 of the 100 most popular VR applications \cite{steam_most_used}.

On conventional platforms, gaming is typically perceived as amongst the most innocuous classes of applications from a security and privacy perspective, while, for example, social media, receives far more attention in this regard. However, recent research indicates that the same may not be true in VR. 
Researchers have already demonstrated the ability to uniquely identify users in VR \cite{miller_personal_2020, nair2023unique} and construct adversarial VR applications that harvest a variety of private user data while being disguised as harmless games \cite{nair2022exploring}.

While the prospect of malicious VR applications poses a legitimate security and privacy threat, most VR games are not deliberately designed to harvest user information. A typical VR game does not include the adversarial challenge and response mechanisms designed to reveal user data in prior work, but does often still broadcast motion data to other players in order to facilitate multi-player functionality.
In this study, we aim to explore the extent to which popular non-adversarial VR games may inadvertently leak private information about their users by revealing their head and hand motion patterns.

It has long been understood that individuals exhibit distinct motion patterns as determined by their unique physiology and muscle memory. Since as early as 1977, studies have demonstrated that these motion patterns can be used not only to uniquely identify individuals \cite{cutting_recognizing_1977}, but also to infer personal characteristics such as age \cite{jain_is_2016} and gender \cite{kozlowski_recognizing_1977, pollick_gender_2005}.
However, the extent to which these findings are applicable to the motion data observable in VR is not yet well understood; although full-body tracking systems are on the horizon, most VR devices today only collect head and hand motion.

To determine whether private information can be inferred from the head and hand motion data broadcast by a typical multi-player VR game, we surveyed players of the popular VR rhythm game ``Beat Saber.'' Players were asked a series of about 50 questions, ranging from demographics like age and gender to personal background, behavioral patterns, and health information. Additionally, participants were asked to provide links to motion-capture recordings of themselves playing Beat Saber on their own personal VR devices.

After collecting data attributes and motion samples from over 1,000 users, we designed 50 binary classification problems based on thresholding the dataset (e.g., ``old'' vs ``young,'' or ``rich'' vs ``poor''). We then trained and tested a deep-learning binary classifier that ingested a sequence of motion data and produced a binary classification for each attribute. We found that over 40 of the 50 attributes could consistently and reliably be inferred from user motion data alone.
Thus, while these users may hold the presumption of anonymity in a VR gaming setting, this presumption is evidently flawed.
Not only are their movement patterns revealing their identity \cite{miller_personal_2020, nair2023unique}, our results imply these patterns could actually be exposing a plethora of information about them to the device, application, server, and even other users within the same virtual environment.

The goal of our work is not to provide an optimal approach for inferring any particular attribute from VR motion data. Rather, we aim to demonstrate, with high statistical significance, that a wide variety of personal and privacy-sensitive variables can be inferred from head and hand motion, in order to highlight the urgent need for privacy-preserving mechanisms in multi-user VR applications.

\medskip

\noindent \textbf{Contributions:}

\vspace{-0.5em}

\begin{itemize}[leftmargin=*]
    \itemsep -0.25em
    \item We surveyed over 1,000 VR users to generate a comprehensive dataset of motion recordings and user data attributes (\S\ref{sec:dataset}).
    \item We present a general-purpose machine learning architecture for inferring user data from head and hand motion streams (\S\ref{sec:method}).
    \item We demonstrate over 40 classes relating to personal user data can be inferred from motion data in standard VR games (\S\ref{sec:evaluation}).
\end{itemize}

\eject

% ml method and procedures

% preview of results

% call to action & significance

%% file: 200-Related-Work.tex
\section{Background and Related Work}
\label{sec:related_work}

This work follows the 2023 Garrido et al. VR privacy systematization of knowledge (SoK) \cite{garrido_sok_2023}, which provides a standard model of VR information flow and threat actors for VR security and privacy research. In this section, we will summarize the threat model of Garrido et al., as well as related work in this area, so as to position our study within the broader landscape of VR privacy research.

\subsection{VR Information Flow and Threat Model}

Most modern consumer-grade VR systems include at least a head-mounted display (HMD) and two hand-held controllers.\footnote{Camera-based hand tracking is an increasingly common alternative.} These systems use an array of sensors to measure the position and orientation of tracked components in 3D space, providing six degrees of freedom (6DoF) per tracked object, typically captured between 60 and 144 times per second, resulting in a ``telemetry stream.''

Many advanced VR devices now contain additional sensors, such as LIDAR arrays, microphones, cameras, full-body tracking, and eye-tracking systems.
However, in this paper, we focus exclusively on the basic telemetry data stream consisting of head and hand motion.
Thus, any information that can be derived from this data stream should be applicable to nearly all VR devices and users.
We expect that the privacy threats discussed in this paper would be exacerbated when motion data is combined with other data streams.

In a typical VR application, the telemetry data stream is forwarded to various parties via the following standard information flow:

\begin{enumerate}[leftmargin=*]
    \itemsep 0em
    \item \textit{Device}. Telemetry data originates at the VR device, which measures tracked objects with 6DoF at between 60 Hz and 144 Hz.
    \item \textit{Application}. A client-side application uses the telemetry data to render a series of auditory, visual, and haptic stimuli, creating an immersive 3D virtual experience for the end user.
    \item \textit{Server}. In the case of a multiplayer or metaverse experience, the application forwards the telemetry stream to an external server.
    \item \textit{Other Users}.  The server, in turn, forwards this data to other users, such that a virtual representation (or ``avatar'') of each user can be rendered on the devices of other users in the same virtual world.
\end{enumerate}

\begin{figure}[h]
\includegraphics[width=0.75 \linewidth]{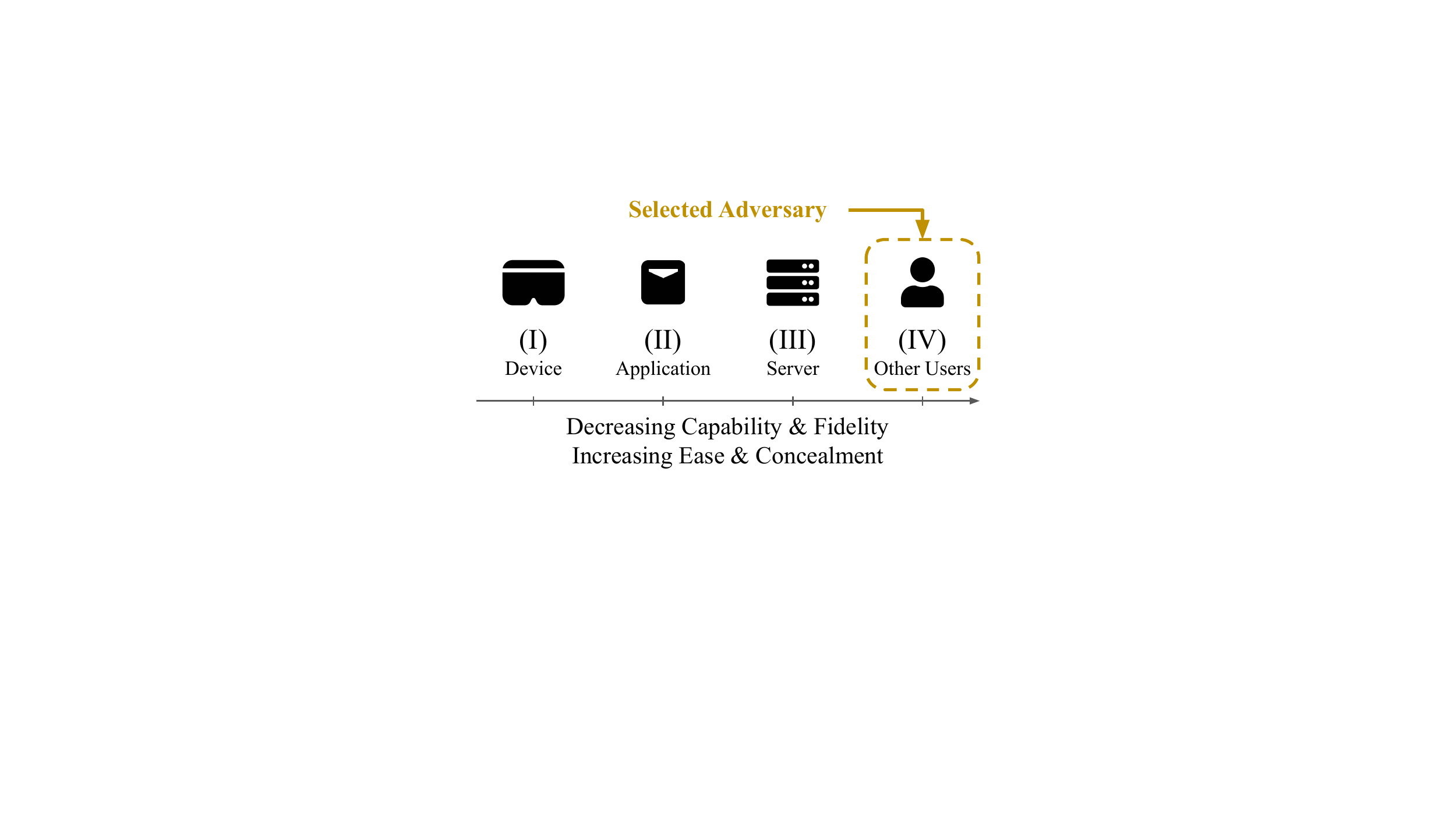}
\centering
\caption{VR threat continuum and selected adversary for this work.}
\label{fig:threat_model}
\end{figure}

Because each entity in the above information flow has access to the telemetry stream of a target user, and could theoretically use it to make adversarial inferences, they are each considered potential adversaries in the Garrido et al. threat model. Adversaries in this model exist on a continuum, as shown in Figure \ref{fig:threat_model}, with adversaries becoming ``weaker'' from left to right due to potential interference, such as compression or transformation, at each step of the data flow.

In this paper, we have chosen to focus on the user adversary (IV) by using only motion data that would normally be available to ordinary users of a multi-user VR application. Because this is considered the weakest adversary in the  Garrido et al. threat model, attacks available to this user can usually also be performed by all other adversaries in the system, while also being amongst the hardest attacks to detect due to their remote and decentralized nature.

\subsection{Human Motion Biometrics}

Since at least the 1970s, researchers have known that individuals reveal identifiable characteristics via their motion. In a 1977 study, Cutting and Kozlowski demonstrated that the gender of 6 individuals could be inferred by a panel of participants using the motion of 8 tracked objects affixed to the body \cite{kozlowski_recognizing_1977}. The study took place before the advent of modern computer graphics, so the authors creatively resorted to taping highly reflective objects to a number of points on the participants' bodies. The researchers then streamed a camera feed of the subjects through a CRT television monitor, and increased the contrast until the participants' silhouettes disappeared and only the individual points of light could be seen. 
Using only the moving points of light visible on the screen, participants were able to infer the gender of the original subjects with up to 63\% accuracy ($p < 0.05$).

More recently, Pollick et al. (2005) \cite{pollick_gender_2005} used statistical features to achieve 79\% accurate identification of gender from motion, while Sarangi et al. (2020) \cite{sarangi2020gender}, and others have used machine learning to achieve inference accuracies of 83\% or more.
Beyond gender, Jain et al. (2016) \cite{jain_is_2016} found that the motion of children can be differentiated from that of adults with 66\% accuracy. Overall, researchers have long known that human motion patterns can be used to infer a variety of personal attributes from human subjects in a laboratory setting.

In one sense, by deriving data from the motion of human body parts tracked in 3D space, these results use data that is highly similar to the motion captured by a VR device. On the other hand, basic VR devices capture only 3 tracked objects, rather than the 8 or more used for motion capture in laboratory studies. Thus, it is not clear whether these previous findings will be transferable to the VR setting.

\subsection{VR Privacy Attacks}
In addition to proposing a standard information flow and threat model, the Garrido et al. SoK tracks a number of existing works in the VR privacy domain. The majority of these works are categorized as ``identification,'' in which VR users are deanonymized or tracked across sessions based on their movement patterns. For instance, Miller et al. (2020) \cite{miller_personal_2020} performed a lab study of 511 users, and then correctly identified users within the pool of 511 with 95\% accuracy using a random forest model. In the largest study to date, Nair et al. (2023) \cite{nair2023unique} used a LightGBM model to identify over 55,000 Beat Saber players with 94.3\% accuracy from their motion. 

A relatively smaller portion of the existing work is categorized as ``profiling,'' whereby specific attributes, such as age or gender, are inferred from users in VR. In one such study, Tricomi et al. (2023) \cite{10027854} accurately infer the gender and age of about 35 VR users, using eye tracking data in addition to head and hand motion.

The remaining studies in this field have focused on the adversarial design of malicious VR applications. Specifically, Aliman and Kester (2020) \cite{9319051} evaluate the risk of generative AI actors harvesting user data in VR, while Nair et al. (2022) \cite{nair2022exploring} present ``MetaData,'' an VR escape room game designed to trick users into revealing personal information via covert adversarially-designed puzzles. In the case of MetaData, the authors were able to recover over 25 personal data attributes, from anthropometrics like height and wingspan to demographics like age and gender, within just a few minutes of gameplay. However, sensors like microphones and cameras were used in addition to motion data to make these inferences.

While the existing literature strongly indicates that VR applications can be used for profiling, it does so mostly by using applications that are explicitly adversarial in design, or by using signals beyond the basic motion data widely available in VR. By contrast, this work aims to demonstrate that profiling is possible even by the weakest known class of adversaries, in popular benign applications like Beat Saber, and from head and hand motion data alone.

% MetaData 
% - adversarial game
% - uses all sensors

% Next, Miller et al. (2020) \cite{pfeuffer_behavioural_2019} conducted a lab study of 511 users, whose telemetry was captured while they watched a series of 360-degree videos in VR. With a random forest model, their system correctly identifies users within the pool of 511 with 95\% accuracy from 5 minutes of telemetry data.

%% file: 300-Dataset.tex
\section{Data Collection}
\label{sec:dataset}

We chose to use ``Beat Saber'' as the model application and main data source for this study, primarily due to the popularity of the game and the relative availability of user data from this application. Furthermore, to maximize ecological validity, we chose to only include participants who were already Beat Saber players.
In this section, we provide background information about Beat Saber and describe our methods for collecting data from existing players.

\subsection{Beat Saber}
``Beat Saber'' \cite{beat_saber} is a VR rhythm game where players slice blocks representing musical beats with a pair of sabers, one held in each hand. With over 6 million copies sold, Beat Saber is the highest-grossing VR application of all time \cite{wobbeking_beat_2022}, and is a representative example of a non-adversarial VR game with multi-player functionality.

\begin{figure}[h]
\includegraphics[width=0.75 \linewidth]{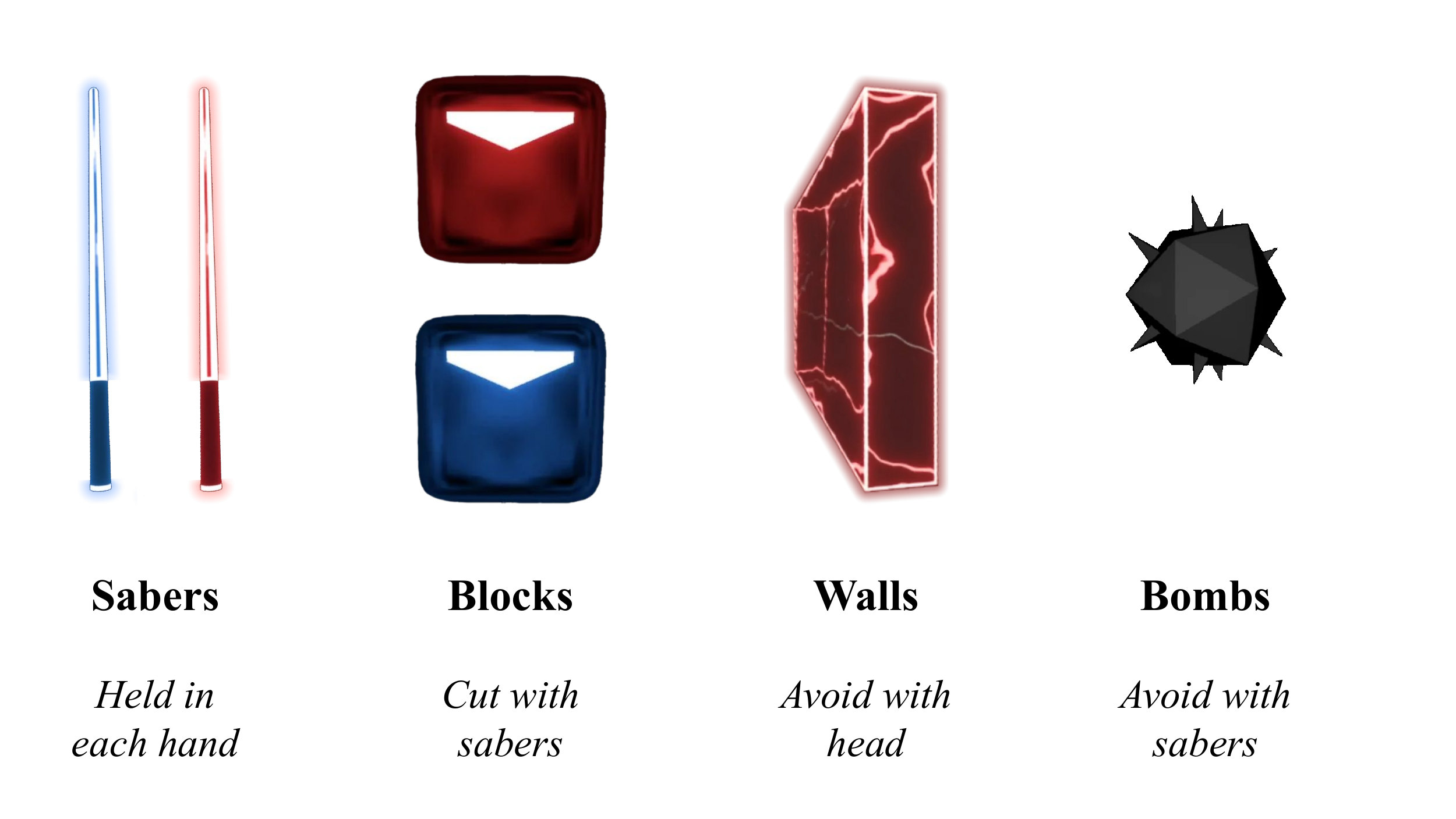}
\centering
\caption{Dynamic objects in ``Beat Saber.''}
\label{fig:beat_saber}
\end{figure}

Beat Saber contains a number of ``maps,'' which consist of an audio track as well as a series of objects presented to the user in time with the audio.
These objects include ``blocks,'' which the player must hit at the correct angle with the correct color saber, ``bombs,'' which the player must avoid hitting with their sabers, and ``walls,'' which the player must avoid with their head (see Fig.~\ref{fig:beat_saber}). At the end of the map, the player is awarded a number of points based on their level of accuracy in completing these tasks. Hundreds of official maps have been added to the game, and over 100,000 user-created maps can be played by installing open-source game modifications.

\subsection{BeatLeader}

``BeatLeader'' \cite{beatleader} is a popular open-source Beat Saber extension and website that offers a third-party leaderboard system for user-created Beat Saber maps.
Beat Saber enthusiasts may choose to install the extension in order to compete with other players to achieve a higher ``rank'' on the leaderboards for popular maps. After playing a Beat Saber map with the BeatLeader extension enabled, scores are automatically uploaded to a globally-visible leaderboard, with over 4 million scores having been uploaded to the platform to date.

When uploading a score to BeatLeader, a recording of the user's motion telemetry during play is automatically captured and attached to their submission. The recording, which utilizes the ``Beat Saber Open Replay'' (BSOR) \cite{bsor} format, is then made publicly available on the BeatLeader website so that it can be used to verify the authenticity of the submitted score.

\subsection{Survey Procedures}
We partnered with the administrators of BeatLeader to conduct an official survey of BeadLeader users, consisting of about 50 personal questions across 9 categories of information. Beat Saber players were invited to take the survey via an announcement released through the official Twitter and Discord accounts of BeatLeader. Participation in the survey was voluntary, with all questions being optional, and no consequences for non-participation. Participants were not monetarily compensated but were given the option to add a unique badge to their BeatLeader profile in recognition of their contribution.

The survey was conducted from April 15th, 2023 to May 1st, 2023, with 1,006 responses collected in that time. The categories of information collected were as follows:

\begin{enumerate}[leftmargin=*]
    \item \textit{Participation}. Participants provided links to their BeatLeader profile from which motion capture recordings could be obtained.
    \item \textit{Demographics}. Participants were asked a variety of demographic questions based on the 2020 United States Census \cite{bureau_2020_nodate}.
    \item \textit{Specifications}. Participants shared an automatically-generated system report containing various computer specifications.
    \item \textit{Background}. Participants were asked about their past history with musical instruments, rhythm games, dancing, and athletics.
    \item \textit{Health}. Participants were asked about their mental and physical health status and disabilities as well as their visual acuity.
    \item \textit{Habits}. Participants were asked about their habits relating to Beat Saber, such as their warmup routine prior to playing.
    \item \textit{Environment}. Participants were asked about the sizes and locations of the areas in which they typically play Beat Saber.
    \item \textit{Anthropometrics}. Participants were asked to measure various physical dimensions of their body, such as height and wingspan.
    \item \textit{Clothing}. Participants were asked about the clothing and footwear they typically wear while playing Beat Saber.
\end{enumerate}

\noindent The exact questions asked in each section are given in Appendix \ref{app:survey}.

\subsection{Motion Recordings}
During the informed consent procedure for the survey, participants also gave us permission to use the publicly-available motion capture recordings from their BeatLeader profile to infer the attributes contained in their survey responses. Accordingly, we downloaded the head and hand motion recordings of each participant from the profile indicated in their survey responses.
For participants with more than 100 recordings, only the latest 100 recordings were utilized.

\subsection{Ethical Considerations}
We conducted this project with significant attention to ethical considerations. Specifically, we refrained from asking questions that could be viewed as overly sensitive, and did not solicit responses from vulnerable populations, including minors under the age of 18. Participants were required to read and agree to a thorough informed consent document prior to inclusion in the study.

An additional source of data was scoring information collected by BeatLeader. This data was already widely publicly available prior to this study, and was available for research in accordance with BeatLeader's privacy policy. Further, participants explicitly agreed to our specific use of this data via the informed consent process.

Participants were not monetarily compensated or given anything of substantial value for their participation in the survey, nor penalized for non-participation. Every question in the survey was optional. Thus, participants were never unduly pressured to provide information that they were uncomfortable with disclosing.

Because the survey responses include sensitive information, such as health status, we followed the strictest data handling standards and guidelines offered by our institution throughout this study.

Overall, we believe this research constitutes a net benefit to society by highlighting the magnitude of the VR privacy threat and motivating future work on defensive countermeasures.

This study has been reviewed and approved by our institution's Institutional Review Board (IRB) as protocol \#2023-03-16120.

%% file: 400-Method.tex
\section{Method}
\label{sec:method}

In this section, we describe our method for determining which of the survey responses are inferrable from VR telemetry data. Specifically, we describe a machine learning model architecture that attempts to infer user data attributes based on a sequential input containing their head and hand motion. Importantly, our goal in this section is not to describe an optimal architecture for inferring any particular attribute, such as age or gender, from motion data. Rather, we aim to describe a general-purpose method for producing binary classifications from VR motion data, and use this method to determine which attributes are present in the motion data with high statistical significance.

\subsection{Binary Classifications}
Our survey results contain a variety of attribute types, including categorical variables such as ethnicity or languages spoken, and numerical variables like height or age, all with different observed distributions. We began by choosing $50$ attributes that we speculated had a reasonable chance of being inferred from motion patterns. To simplify our analysis, we then turned each of these attributes into a binary classification problem. For example, marital status was turned into a binary classification of ``never married'' versus all other responses (married, divorced, etc.). For continuous attributes, such as height, a wide rejection band was usually incorporated. The exact binary splits for all attributes are given in Appendix \ref{app:dist}.

Using this approach allows us to use a single binary classification model architecture and statistical analysis technique for all attributes being considered. This simplified approach is sufficient for our purposes of demonstrating whether the attribute can be inferred from VR motion data, though regression or multi-class classification models may be more suitable for use in a real-world deployment.\footnote{We stress that deployment in a context where the user has not specifically and knowingly agreed to this type of monitoring would raise significant ethical concerns, particularly if the data remained linked to the user's identity.}

\subsection{Model Architecture}
We evaluated the efficacy of a variety of machine learning architectures, including Random Forest, CNN, LSTM, and Transformer models, for performing our binary classification task using the sequential motion data. We found the Transformer-based models to be most effective at inferring a majority of the chosen attributes.

The Transformer model \cite{10.5555/3295222.3295349} incorporates a self-attention mechanism to capture dependencies and relationships within an input sequence. Unlike other deep learning models that process the elements of a sequence sequentially, the Transformer simultaneously processes all elements in parallel, allowing it to weigh the importance of each element in the context of the whole sequence.

\begin{figure}[h]
\includegraphics[width=\linewidth]{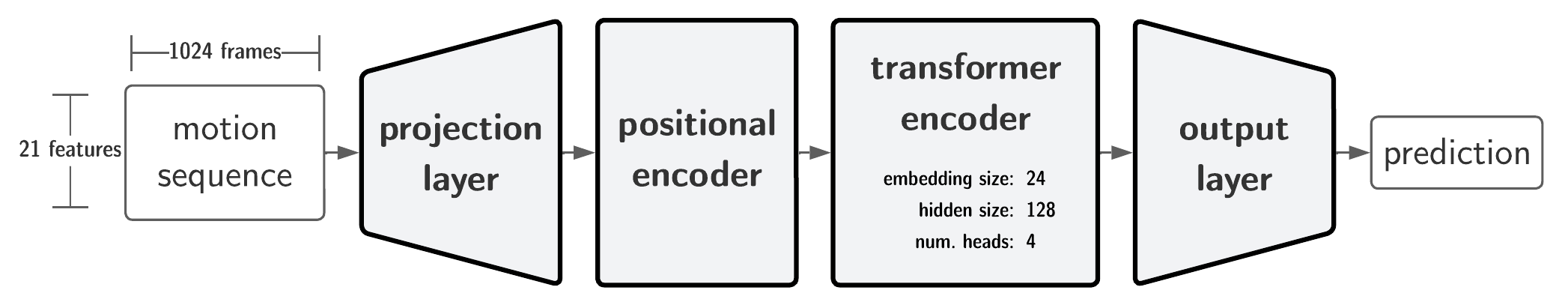}
\centering
\caption{Transformer model architecture.}
\label{fig:arch}
\end{figure}

FIG. \ref{fig:arch} illustrates our Transformer implementation. Input sequences first go through a projection layer that prepares the features by increasing the dimensionality to the embedding size of the Transformer. Following this, the data pass an encoding layer that applies a sinusoidal positional encoding, which adds information about the relative position of each element in the input sequence. This step is important, as Transformers do not have an inherent notion of order or position. Next, we use the Transformer encoder component to generate a contextualized representation of the input sequence. Finally, a fully connected output layer reduces the encoder output to a scalar value, which provides the binary classification.

% For example, in the Age attribute, the following accuracy values were observed across various architectures with the same input features:

% \begin{itemize}[leftmargin=*]
%     \itemsep 0em
%     \item Transformers: \textbf{85.0\%}
%     \item LSTM: 75.0\%
%     \item LightGBM: 75.0\%
%     \item CNN: 65.0\%
%     \item Random Forest: \textbf{65.0\%}
% \end{itemize}

\subsection{Model Input}
\label{sec:featurization}
An advantage of transformer models is that they are intrinsically well-suited for handling time-series data.
We thus chose to use a sequential featurization method to encode the motion of VR users in 3D space over a period of time. At a given time step (``frame''), we capture the position and orientation of three objects (the user's head and two hands) in 3D space. Each tracked object is captured using three positional coordinates and four orientation coordinates expressed as a quaternion. In total, $7$ coordinates are taken for each object, resulting in a total of $21$ values captured per frame. For each motion recording, we sample the first 1,024 frames to provide as input to our model. Thus, any given recording is represented by a $(21\times1024)$-dimensional input; recordings with less than 1024 frames were zero-padded. The frames were sampled at their original frame rate without interpolation or normalization, as the model's normalization layer already allows it to rescale inputs internally.

% Our best-performing model is a Transformer model based a custom architecture with the following hyperparameters:

% \begin{itemize}[leftmargin=*]
%     \itemsep 0em
%     \item Input Shape: $(1024\times21)$
%     \item Embedding Size: 24
%     \item Hidden Size: 128
%     \item Number of Layers: 2
%     \item Output Size: 1
%     \item Learning Rate: 0.00002
%     \item Epochs: 100
%     \item Batch Size: 32
% \end{itemize}

\subsection{Data Split}

Using the BeatLeader database, we downloaded the 100 most recent motion recordings from each of the users who responded to our survey. We then converted each recording into a $(21\times1024)$-dimensional input using the featurization method of \S\ref{sec:featurization}.

For each of the 50 attributes, we selected 20 users from each of the two classes to split between testing and validation sets, with the remaining users being used for training. We then resampled the training sequences to select $10,000$ recordings for training each class. As such, all three sets were perfectly balanced between both classes of every binary attribute, with $10,000$ recordings for training each class and $1,000$ recordings for validating and testing each class. This process was repeated across 3 to 7 Monte Carlo cross-validations \cite{xu2001monte} for each attribute to assess statistical significance.

\subsection{Training}

We evaluated the machine learning approach by using PyTorch v2.0.1 to train and test one binary classification model for each of the 50 selected attributes. We utilized the Adam optimization algorithm \cite{kingma2017adam} with a binary cross-entropy (BCE) loss function. Each model was trained across 100 epochs, with the best-performing epoch then being selected using a validation set.
The evaluation was performed using a single machine with an AMD Ryzen 9 5950X 16-core CPU, 128 GB of DDR4 RAM, and an NVIDIA GeForce RTX 3090 GPU. With this setup, each model took an average of 37 minutes to train and test, with the entire evaluation taking approximately 31 hours.

\subsection{Evaluation Metrics}
\label{sec:metrics}

Because our sampling technique always includes the same number of sequences and users in each class, the statistical significance of these results can be evaluated using a cumulative binomial test where $n$ is the number of samples, $K$ is the number of correct predictions, and $p_{\emptyset}$ is $0.5$. We use this as our primary target metric in \S\ref{sec:evaluation}.
The use of completely balanced training, testing, and validation sets substantially diminishes the need for more nuanced statistical tests, such as the $F_1$-score \cite{sasaki2007truth} or Cohen's kappa \cite{cohen1960coefficient}.

\subsection{Hyperparameter Tuning}

We performed a tuning sweep of the relevant hyperparameters (hidden size, learning rate, etc.) using just two attributes, \verb|StandaloneGrip| and \verb|Sex|.
The hyperparameters that maximized the significance of these attributes per the metrics in \S\ref{sec:metrics} were then used throughout our evaluation and are provided in Appendix \ref{app:params}.

%% file: 510-TAB-Results.tex
% \usepackage{graphicx}
\begin{table}[]
\resizebox{\columnwidth}{!}{%
\begin{tabular}{l|cccc|cccc|}
\cline{2-9}
 & \multicolumn{4}{c|}{\textbf{Per Sequence}} & \multicolumn{4}{c|}{\textbf{Per User}} \\ \hline
\multicolumn{1}{|l|}{\textbf{Attribute}} & \textbf{Total \#} & \textbf{Test \#} & \textbf{Accuracy} & \textbf{Significance} & \textbf{Total \#} & \textbf{Test \#} & \textbf{Accuracy} & \textbf{Significance} \\ \hline
\multicolumn{1}{|l|}{StandaloneGrip} & 31,100 & 6,000 & 85.9\% & \cellcolor[HTML]{93C47D}p \textless 0.001 & 311 & 60 & 91.7\% & \cellcolor[HTML]{93C47D}p \textless 0.001 \\ \hline
\multicolumn{1}{|l|}{Height} & 19,100 & 6,000 & 76.5\% & \cellcolor[HTML]{93C47D}p \textless 0.001 & 191 & 60 & 86.7\% & \cellcolor[HTML]{93C47D}p \textless 0.001 \\ \hline
\multicolumn{1}{|l|}{Controller} & 33,200 & 6,000 & 81.2\% & \cellcolor[HTML]{93C47D}p \textless 0.001 & 332 & 60 & 85.0\% & \cellcolor[HTML]{93C47D}p \textless 0.001 \\ \hline
\multicolumn{1}{|l|}{Weight} & 9,800 & 6,000 & 73.6\% & \cellcolor[HTML]{93C47D}p \textless 0.001 & 98 & 60 & 85.0\% & \cellcolor[HTML]{93C47D}p \textless 0.001 \\ \hline
\multicolumn{1}{|l|}{FootSize} & 9,100 & 6,000 & 73.2\% & \cellcolor[HTML]{93C47D}p \textless 0.001 & 91 & 60 & 85.0\% & \cellcolor[HTML]{93C47D}p \textless 0.001 \\ \hline
\multicolumn{1}{|l|}{Country} & 33,300 & 6,000 & 60.3\% & \cellcolor[HTML]{93C47D}p \textless 0.001 & 333 & 60 & 81.7\% & \cellcolor[HTML]{93C47D}p \textless 0.001 \\ \hline
\multicolumn{1}{|l|}{RhythmGames} & 10,900 & 6,000 & 63.5\% & \cellcolor[HTML]{93C47D}p \textless 0.001 & 109 & 60 & 80.0\% & \cellcolor[HTML]{93C47D}p \textless 0.001 \\ \hline
\multicolumn{1}{|l|}{Age} & 62,300 & 6,000 & 64.9\% & \cellcolor[HTML]{93C47D}p \textless 0.001 & 623 & 60 & 78.3\% & \cellcolor[HTML]{93C47D}p \textless 0.001 \\ \hline
\multicolumn{1}{|l|}{TotalPlayTime} & 34,400 & 6,000 & 67.7\% & \cellcolor[HTML]{93C47D}p \textless 0.001 & 344 & 60 & 78.3\% & \cellcolor[HTML]{93C47D}p \textless 0.001 \\ \hline
\multicolumn{1}{|l|}{Headset} & 65,000 & 6,000 & 66.9\% & \cellcolor[HTML]{93C47D}p \textless 0.001 & 650 & 60 & 76.7\% & \cellcolor[HTML]{93C47D}p \textless 0.001 \\ \hline
\multicolumn{1}{|l|}{LeftArm} & 10,300 & 6,000 & 65.2\% & \cellcolor[HTML]{93C47D}p \textless 0.001 & 103 & 60 & 76.7\% & \cellcolor[HTML]{93C47D}p \textless 0.001 \\ \hline
\multicolumn{1}{|l|}{RightArm} & 10,200 & 6,000 & 64.9\% & \cellcolor[HTML]{93C47D}p \textless 0.001 & 102 & 60 & 75.0\% & \cellcolor[HTML]{93C47D}p \textless 0.001 \\ \hline
\multicolumn{1}{|l|}{Athletics} & 8,700 & 6,000 & 59.1\% & \cellcolor[HTML]{93C47D}p \textless 0.001 & 87 & 60 & 75.0\% & \cellcolor[HTML]{93C47D}p \textless 0.001 \\ \hline
\multicolumn{1}{|l|}{MaritalStatus} & 81,400 & 6,000 & 60.2\% & \cellcolor[HTML]{93C47D}p \textless 0.001 & 814 & 60 & 73.3\% & \cellcolor[HTML]{93C47D}p \textless 0.001 \\ \hline
\multicolumn{1}{|l|}{EmploymentStatus} & 64,200 & 6,000 & 65.1\% & \cellcolor[HTML]{93C47D}p \textless 0.001 & 642 & 60 & 71.7\% & \cellcolor[HTML]{93C47D}p \textless 0.001 \\ \hline
\multicolumn{1}{|l|}{AnyRhythmGames} & 83,000 & 6,000 & 54.8\% & \cellcolor[HTML]{93C47D}p \textless 0.001 & 830 & 60 & 70.0\% & \cellcolor[HTML]{93C47D}p \textless 0.001 \\ \hline
\multicolumn{1}{|l|}{Ethnicity} & 73,900 & 6,000 & 59.7\% & \cellcolor[HTML]{93C47D}p \textless 0.001 & 739 & 60 & 70.0\% & \cellcolor[HTML]{93C47D}p \textless 0.001 \\ \hline
\multicolumn{1}{|l|}{SteamComputerFormFactor} & 51,300 & 6,000 & 58.5\% & \cellcolor[HTML]{93C47D}p \textless 0.001 & 513 & 60 & 70.0\% & \cellcolor[HTML]{93C47D}p \textless 0.001 \\ \hline
\multicolumn{1}{|l|}{Footwear} & 36,700 & 6,000 & 60.5\% & \cellcolor[HTML]{93C47D}p \textless 0.001 & 367 & 60 & 70.0\% & \cellcolor[HTML]{93C47D}p \textless 0.001 \\ \hline
\multicolumn{1}{|l|}{AnyVRRhythmGames} & 83,000 & 8,000 & 56.8\% & \cellcolor[HTML]{93C47D}p \textless 0.001 & 830 & 80 & 68.8\% & \cellcolor[HTML]{93C47D}p \textless 0.001 \\ \hline
\multicolumn{1}{|l|}{Income} & 76,700 & 8,000 & 55.0\% & \cellcolor[HTML]{93C47D}p \textless 0.001 & 767 & 80 & 68.8\% & \cellcolor[HTML]{93C47D}p \textless 0.001 \\ \hline
\multicolumn{1}{|l|}{Wingspan} & 16,000 & 8,000 & 59.9\% & \cellcolor[HTML]{93C47D}p \textless 0.001 & 160 & 80 & 68.8\% & \cellcolor[HTML]{93C47D}p \textless 0.001 \\ \hline
\multicolumn{1}{|l|}{Handedness} & 71,600 & 10,000 & 55.2\% & \cellcolor[HTML]{93C47D}p \textless 0.001 & 716 & 100 & 66.0\% & \cellcolor[HTML]{93C47D}p \textless 0.001 \\ \hline
\multicolumn{1}{|l|}{HandLength} & 51,000 & 8,000 & 58.5\% & \cellcolor[HTML]{93C47D}p \textless 0.001 & 510 & 80 & 66.3\% & \cellcolor[HTML]{B6D7A8}p = 0.002 \\ \hline
\multicolumn{1}{|l|}{SubstanceUse} & 69,200 & 10,000 & 55.9\% & \cellcolor[HTML]{93C47D}p \textless 0.001 & 692 & 100 & 64.0\% & \cellcolor[HTML]{B6D7A8}p = 0.002 \\ \hline
\multicolumn{1}{|l|}{Preparation} & 39,400 & 8,000 & 58.2\% & \cellcolor[HTML]{93C47D}p \textless 0.001 & 394 & 80 & 65.0\% & \cellcolor[HTML]{B6D7A8}p = 0.005 \\ \hline
\multicolumn{1}{|l|}{LowerBody} & 29,500 & 8,000 & 55.9\% & \cellcolor[HTML]{93C47D}p \textless 0.001 & 295 & 80 & 65.0\% & \cellcolor[HTML]{B6D7A8}p = 0.005 \\ \hline
\multicolumn{1}{|l|}{Lenses} & 80,900 & 8,000 & 55.3\% & \cellcolor[HTML]{93C47D}p \textless 0.001 & 809 & 80 & 65.0\% & \cellcolor[HTML]{B6D7A8}p = 0.005 \\ \hline
\multicolumn{1}{|l|}{Languages} & 80,700 & 8,000 & 56.5\% & \cellcolor[HTML]{93C47D}p \textless 0.001 & 807 & 80 & 65.0\% & \cellcolor[HTML]{B6D7A8}p = 0.005 \\ \hline
\multicolumn{1}{|l|}{SteamOperatingSystemVersion} & 50,800 & 8,000 & 58.4\% & \cellcolor[HTML]{93C47D}p \textless 0.001 & 508 & 80 & 65.0\% & \cellcolor[HTML]{B6D7A8}p = 0.005 \\ \hline
\multicolumn{1}{|l|}{Music} & 29,600 & 8,000 & 53.6\% & \cellcolor[HTML]{93C47D}p \textless 0.001 & 296 & 80 & 65.0\% & \cellcolor[HTML]{B6D7A8}p = 0.005 \\ \hline
\multicolumn{1}{|l|}{AnyMentalDisabilities} & 83,000 & 10,000 & 52.6\% & \cellcolor[HTML]{93C47D}p \textless 0.001 & 830 & 100 & 63.0\% & \cellcolor[HTML]{B6D7A8}p = 0.006 \\ \hline
\multicolumn{1}{|l|}{Sex} & 76,300 & 10,000 & 56.5\% & \cellcolor[HTML]{93C47D}p \textless 0.001 & 763 & 100 & 63.0\% & \cellcolor[HTML]{B6D7A8}p = 0.006 \\ \hline
\multicolumn{1}{|l|}{AnyPhysicalDisabilities} & 83,000 & 10,000 & 54.5\% & \cellcolor[HTML]{93C47D}p \textless 0.001 & 830 & 100 & 62.0\% & \cellcolor[HTML]{D9EAD3}p = 0.010 \\ \hline
\multicolumn{1}{|l|}{ReactionTime} & 9,800 & 14,000 & 53.1\% & \cellcolor[HTML]{93C47D}p \textless 0.001 & 98 & 140 & 60.0\% & \cellcolor[HTML]{D9EAD3}p = 0.011 \\ \hline
\multicolumn{1}{|l|}{AnyMusic} & 83,000 & 8,000 & 55.7\% & \cellcolor[HTML]{93C47D}p \textless 0.001 & 830 & 80 & 62.5\% & \cellcolor[HTML]{D9EAD3}p = 0.016 \\ \hline
\multicolumn{1}{|l|}{AnyAthletics} & 19,900 & 8,000 & 55.7\% & \cellcolor[HTML]{93C47D}p \textless 0.001 & 199 & 80 & 61.3\% & \cellcolor[HTML]{D9EAD3}p = 0.016 \\ \hline
\multicolumn{1}{|l|}{EducationalStatus} & 62,200 & 8,000 & 57.1\% & \cellcolor[HTML]{93C47D}p \textless 0.001 & 622 & 80 & 60.0\% & \cellcolor[HTML]{D9EAD3}p = 0.028 \\ \hline
\multicolumn{1}{|l|}{IPD} & 6,700 & 8,000 & 55.8\% & \cellcolor[HTML]{93C47D}p \textless 0.001 & 67 & 80 & 60.0\% & \cellcolor[HTML]{D9EAD3}p = 0.028 \\ \hline
\multicolumn{1}{|l|}{Dance} & 82,000 & 10,000 & 52.3\% & \cellcolor[HTML]{93C47D}p \textless 0.001 & 820 & 100 & 59.0\% & \cellcolor[HTML]{D9EAD3}p = 0.028 \\ \hline
\multicolumn{1}{|l|}{PoliticalOrientation} & 33,100 & 10,000 & 53.5\% & \cellcolor[HTML]{93C47D}p \textless 0.001 & 331 & 100 & 58.0\% & \cellcolor[HTML]{D9EAD3}p = 0.044 \\ \hline
\multicolumn{1}{|l|}{UpperBody} & 47,200 & 10,000 & 52.0\% & \cellcolor[HTML]{93C47D}p \textless 0.001 & 472 & 100 & 57.0\% & p = 0.067 \\ \hline
\multicolumn{1}{|l|}{SteamProcessorLogicalCores} & 33,500 & 10,000 & 51.0\% & \cellcolor[HTML]{D9EAD3}p = 0.021 & 335 & 100 & 56.0\% & p = 0.136 \\ \hline
\multicolumn{1}{|l|}{HadCOVID} & 83,000 & 10,000 & 54.4\% & \cellcolor[HTML]{93C47D}p \textless 0.001 & 830 & 100 & 55.0\% & p = 0.136 \\ \hline
\multicolumn{1}{|l|}{CaffinatedBeverages} & 40,800 & 10,000 & 52.9\% & \cellcolor[HTML]{93C47D}p \textless 0.001 & 408 & 100 & 55.0\% & p = 0.136 \\ \hline
\multicolumn{1}{|l|}{RoomArea} & 33,100 & 8,000 & 50.5\% & p = 0.183 & 331 & 80 & 56.3\% & p = 0.157 \\ \hline
\multicolumn{1}{|l|}{PhysicalFitness} & 7,800 & 12,000 & 54.2\% & \cellcolor[HTML]{93C47D}p \textless 0.001 & 78 & 120 & 55.0\% & p = 0.158 \\ \hline
\multicolumn{1}{|l|}{SteamProcessorCPUVendor} & 51,600 & 10,000 & 49.2\% & p = 0.953 & 516 & 100 & 53.0\% & p = 0.309 \\ \hline
\multicolumn{1}{|l|}{SteamLighthouses} & 5,500 & 8,000 & 48.6\% & p = 0.993 & 55 & 80 & 52.5\% & p = 0.369 \\ \hline
\multicolumn{1}{|l|}{ColorBlindness} & 79,800 & 10,000 & 50.4\% & p = 0.227 & 798 & 100 & 52.0\% & p = 0.382 \\ \hline
\end{tabular}%
}
\vspace{1em}
\caption{Accuracy of inferring 50 attributes from head and hand motion data, with statistical significance calculated via binomial tests.}
\label{tab:results}
\end{table}

%% file: 500-Evaluation.tex
\section{Results}
\label{sec:evaluation}

% results
% p-value with binomial test
% # significant
% # non-significant

% \subsection{Results}

After training a model for each of the tested attributes, we first generated a classification for all of the 100 sequences per user for every user in the testing set. Next, we generated a prediction for each user by taking the average classification across their 100 sequences. Table \ref{tab:results} shows the accuracy of the results per sequence and per, along with the p-values corresponding to the metrics described in \S\ref{sec:metrics}.

Overall, we observe that 33 of the 50 attributes were predicted with high statistical significance ($p < 0.01$), and another 8 of 50 with moderate statistical significance ($p < 0.05$) on a per-user basis. On a per-sequence basis, 45 out of 50 attributes were highly significant ($p < 0.01$), and one was moderately significant ($p < 0.05$). The difference in significance is largely accounted for by sample size; in total, 100 times more recordings were present than users.

\subsection{Macro Significance}

Given that we evaluated $50$ attributes in this work, only a portion of which were inferred with significant accuracy, it remains to be demonstrated that the overall evaluation was statistically significant.
To assess the overall significance of our result, we performed a secondary evaluation in which the trained models from our main evaluation were tested with randomly-generated fictitious inputs.
We then performed a Wilcoxon signed-rank test to compare the distribution of classification accuracy values across the 50 attributes on these fictitious inputs with the distribution of true results in Table~\ref{tab:results}. 
We found $p < 0.0001$ on both a per-sequence and per-user basis, indicating a high overall statistical significance of our results.

%% file: 600-Discussion.tex
\section{Discussion}
\label{sec:discussion}

Though the threat of adversarial game design in VR \cite{9319051, nair2022exploring} remains salient, in this study, we have shown that even a seemingly harmless VR rhythm game reveals enough motion data to infer a wide variety of user characteristics. 
``Beat Saber'' is not particularly conducive to data harvesting, with a simple ruleset and interaction model. For instance, there are no in-game characters to interact with, which could reveal even more information than we already observed.

In comparison with prior work, the setting evaluated in this paper represents a realistic and challenging threat scenario.
Unlike laboratory studies, which take advantage of a controlled environment with homogeneous hardware and firmware, our data comes from real VR users around the world with a wide variety of devices and environments. We further limited our models to only use head and hand motion data, discarding other data modalities used in prior work, and used the weakest adversary class for our evaluations.

Despite the inherent difficulty of this dataset, a large number of personal attributes were accurately and consistently inferrable from the XR motion data alone. These attributes go beyond the obvious anthropometric measurements to include a surprising amount of information about the player's background, demographics, environment, habits, and even health.
Many of these attributes, such as disability status, could be considered highly private information by end users.
Others, like political orientation, have historically been used by social media platforms for targeted advertising and could similarly leave VR users open to targeted influence campaigns. 

There are also a number of avenues adversaries could pursue to further improve VR profiling capabilities. Since VR motion patterns are now known to constitute uniquely identifiable biometrics \cite{nair2023unique}, adversaries are not limited to collecting data from a single application. Rather, they could leverage the identifiability of VR users to track them across applications and usage sessions, building a rich user profile over time. These risks are further exacerbated with the introduction of data from additional sensors, such as microphones, cameras, LIDAR arrays, and eye and body tracking, all of which may provide data beyond the head and hand motion considered herein.

Despite using terms like ``attack'' and ``adversary'' throughout this paper, there is nothing inherently unlawful about collecting data from VR users, who may agree to such data harvesting in terms of service agreements.
However, while users have developed a level of understanding about data collection on the web, most lack awareness of the breadth of personal information that can be extracted from even the simplest of VR experiences, such as a rhythm game.

As it stands, major VR device manufacturers have been observed selling hardware at a loss of up to \$10 billion per year \cite{published_oculus_2021}.
Given the deep roots of major metaverse players in the advertising industry, there could be at least a temptation to leverage existing sales channels to monetize the rich user data inferable from motion in VR. Our results indicate were they to act on this temptation, VR manufacturers and developers could reasonably harvest a vast array of user information from motion data collected in VR applications, including sensitive data about their age, gender, weight, ethnicity, income, substance use, disability status, and more.

\subsection{Limitations}

The motion recordings used in this study originate entirely from a single game. While ``Beat Saber'' is the most popular VR game to date \cite{wobbeking_beat_2022}, and is a representative example of a non-adversarial VR game, we cannot yet demonstrate that our findings will generalize to other types of VR applications. Furthermore, we chose to only survey existing Beat Saber players, and are unsure whether novice players, who would potentially demonstrate less consistent movement patterns, would be equally susceptible to these inferences.

We used the game recordings to infer a series of attributes that were self-reported via an online survey, and were thus subject to the biases typically associated with self-reported data. The participants in this survey were also not representative of the general population; for example, over 80\% of respondents were male. However, the sample is fairly representative of the current VR user population \cite{noauthor_report_2017}. The distribution of each attribute is given in Appendix \ref{app:dist}. Unbalanced distributions did not inflate the reported results, as each binary class was rebalanced prior to training and testing.

Finally, a portion of the reported findings may be the result of hidden correlations rather than direct inference.
For example, it is likely that some attributes like employment or marital status are not directly observable from motion, but are correlated to age, which can be inferred from motion data.
These correlations could apply generally to human motion, but may also represent sampling or response biases.
Appendix \ref{app:corr} shows the correlation between each pair of attributes. 
Due to the difficulty of explaining the internal function of deep learning models, we cannot easily determine the mechanisms of causality associated with each result. However, we consider the potential to infer this data from VR users to be noteworthy and concerning, regardless of the cause.

\subsection{Future Work}

Our major motivation for conducting this study is to highlight the need for future research into defensive countermeasures for safeguarding user motion data in VR.
Unfortunately, doing so is non-trivial, as sharing motion data is genuinely required for a variety of legitimate purposes in VR applications. Furthermore, motion patterns are largely subconscious and are often deeply ingrained in muscle memory, and are thus hard to intentionally obscure.

A first step towards defending against motion-based inference attacks is gaining a better understanding of how they work. While deep learning models are notoriously difficult to explain, we hope to see future work that uses advanced model explainability techniques to better understand the mechanisms underlying our results.

Researchers have already evaluated the use of differential privacy in VR \cite{nair2022going}, with moderate success at obscuring anthropometric attributes like height and wingspan. However, it is unclear whether this approach would be effective at defeating the sequential machine learning models used to derive more complex attributes in this paper.

Future work should evaluate the use of ``corruption models'' designed to obscure sensitive attributes embedded in VR motion data while minimally impacting legitimate application functionality.
Generative adversarial networks (GAN) have already been shown to be effective at hiding attributes like gender from sequential data \cite{gan}, and could likely be applied to VR telemetry data streams as well.

Another defense worth exploring is the use of trusted execution environments (TEEs) to provide auditability for metaverse servers that utilize telemetry data. TEEs like Intel's SGX could provide a hardware-based attestation mechanism that allows users to verify that servers are only using their motion data for legitimate purposes.

%% file: 700-Conclusion.tex
\section{Conclusion}
\label{sec:conclusion}

With major new products like the Apple Vision Pro on the horizon, extended reality technologies are on track to soon become a ubiquitous means of accessing the internet. For the foreseeable future, motion tracking ``telemetry'' data will remain at the core of nearly all extended reality and metaverse experiences. In this study, we demonstrated that even in non-adversarial applications, this data stream carries significant privacy implications for VR users.

Privacy risks are not unique to VR, and users have grown accustomed to some level of data harvesting on most internet platforms.
However, unlike in conventional web applications, users are largely unaware of the unique privacy risks associated with VR applications, and lack the suite of defensive tools, such as privacy-preserving browser extensions, that have been developed over time for the web.

Thus, we are currently at a crossroads. If nothing is done to improve the present security and privacy posture of virtual reality, it is poised to inherit an exaggerated version of the privacy issues that are prevalent on the web. However, we can also take the opportunity to learn from the history of browser-based attacks and defenses. We hope the results of this study motivate security and privacy practitioners to prioritize research in this field and build privacy-preserving mechanisms into the fabric of future metaverse platforms.

% \eject

%% file: 900-Survey.tex
\newcommand\nothing[1]{}

\subsection{Participation}

\noindent \textbf{Mods.} Have you ever played BeatSaber with the ScoreSaber and/or BeatLeader mods installed?
\nothing{img/dist/Mods.pdf}

\noindent \textbf{Secondary Accounts.} Have you ever submitted a score using a BeatLeader or ScoreSaber account not listed above?
\nothing{img/dist/HasSecondary.pdf}

\noindent \textbf{Multiple Users.} Has any person other than yourself ever submitted a score to any of the BeatLeader or ScoreSaber accounts listed above?
\nothing{img/dist/MultipleUsers.pdf}

\noindent \textbf{Play Time.} To the nearest hour, how many total hours have you spent playing Beat Saber?
\nothing{img/dist/TotalPlayTime.pdf}

\subsection{Demographics}

\noindent \textbf{Sex.} What is your sex?
\nothing{img/dist/Sex.pdf}

\noindent \textbf{Age.} What is your age in years?
\nothing{img/dist/Age.pdf}

\noindent \textbf{Employment Status.} Which of the following options best represents your current employment status?
\nothing{img/dist/EmploymentStatus.pdf}

\noindent \textbf{Marital Status.} Which of the following options best represents your current marital status?
\nothing{img/dist/MaritalStatus.pdf}

\noindent \textbf{Languages.} Which languages do you speak fluently? If multiple, list all languages spoken in order of proficiency.
\nothing{img/dist/Languages.pdf}

\noindent \textbf{Educational Status.} What is the highest degree or level of school you have completed?
\nothing{img/dist/EducationalStatus.pdf}

\noindent \textbf{Income.} Which of the following options best represents your total gross income in 2022? Convert your answer to United States Dollars (USD).
\nothing{img/dist/Income.pdf}

\noindent \textbf{Ethnicity.} What is your ethnicity?
\nothing{img/dist/Ethnicity.pdf}

\noindent \textbf{Political Orientation.} Which of the following generally best represents your political views?
\nothing{img/dist/PoliticalOrientation.pdf}

\subsection{Specifications}

\noindent \textbf{CPU Brand.} According to the Steam system report, what is the vendor of the CPU in the user's PC?
\nothing{img/dist/SteamProcessorCPUVendor.pdf}

\noindent \textbf{Logical Cores.} According to the Steam system report, how many logical CPU cores are in the user's PC?
\nothing{img/dist/SteamProcessorLogicalCores.pdf}

\noindent \textbf{CPU Speed.} According to the Steam system report, what is the base CPU clock speed in the user's PC?
\nothing{img/dist/SteamProcessorSpeed.pdf}

\noindent \textbf{Form Factor.} According to the Steam system report, is the user's PC a laptop or desktop?
\nothing{img/dist/SteamComputerFormFactor.pdf}

\noindent \textbf{Operating System.} According to the Steam system report, what is the operating system of the user's PC?
\nothing{img/dist/SteamOperatingSystemVersion.pdf}

\noindent \textbf{System Memory.} According to the Steam system report, how much RAM is in the user's PC?
\nothing{img/dist/SteamMemoryRAM.pdf}

\noindent \textbf{Drive Space.} According to the Steam system report, how much empty disk space is in the user's PC?
\nothing{img/dist/SteamMiscellaneousTotalHardDiskSpaceAvailable.pdf}

\noindent \textbf{Base Stations.} According to the Steam system report, how many lighthouses or base stations does the user have?
\nothing{img/dist/SteamLighthouses.pdf}

\noindent \textbf{Graphics Card.} According to the Steam system report, what is the primary GPU of the user's PC?
\nothing{img/dist/SteamVideoCardDriver.pdf}

\subsection{Background}

\noindent \textbf{Music.} Have you ever skillfully played a musical instrument?
\nothing{img/dist/AnyMusic.pdf}

\noindent \textbf{Music.} If you have ever skillfully played a musical instrument, list the instrument(s).
\nothing{img/dist/Music.pdf}

\noindent \textbf{Dance.} Have you ever skillfully practiced or exhibited a recognized form of dance?
\nothing{img/dist/Dance.pdf}

\noindent \textbf{Rhythm Games.} Have you ever played a rhythm game other than Beat Saber?
\nothing{img/dist/AnyRhythmGames.pdf}

\noindent \textbf{Rhythm Games.} If you have ever played a rhythm game other than Beat Saber, list the game(s).
\nothing{img/dist/RhythmGames.pdf}

\noindent \textbf{Athletics.} Have you ever competitively participated in an individual or team-based athletic sport?
\nothing{img/dist/AnyAthletics.pdf}

\noindent \textbf{Athletics.} If you have ever competitively participated in an individual or team-based athletic sport, list the sport(s).
\nothing{img/dist/Athletics.pdf}

\subsection{Health}

\noindent \textbf{Eyesight.} Do you regularly wear prescription glasses or contact lenses?
\nothing{img/dist/Eyesight.pdf}

\noindent \textbf{Lenses.} Do you usually wear prescription  glasses or contact lenses while playing Beat Saber?
\nothing{img/dist/Lenses.pdf}

\noindent \textbf{Color Blindness.} Do you have any form of color blindness or color vision deficiency?
\nothing{img/dist/ColorBlindness.pdf}

\noindent \textbf{Physical Disabilities.} Have you ever been diagnosed with a physical disability or other physical health condition?
\nothing{img/dist/AnyPhysicalDisabilities.pdf}

\noindent \textbf{Mental Disabilities.} Have you ever been diagnosed with a mental disability or other mental health condition?
\nothing{img/dist/AnyMentalDisabilities.pdf}

\noindent \textbf{Illness.} In the past year, have you experienced COVID-19?
\nothing{img/dist/HadCOVID.pdf}

\subsection{Habits}

\nothing{img/grip.png}

\noindent \textbf{Grip.} Which of the following grips do you prefer to use on standalone VR devices (e.g., Oculus Quest 2, PICO Neo3, etc.)?
\nothing{img/dist/StandaloneGrip.pdf}

\noindent \textbf{Preparation.} Which of the following activities, if any, do you perform immediately before playing Beat Saber? Select all that apply.
\nothing{img/dist/Preparation.pdf}

\noindent \textbf{Physical Fitness.} How would you rate your current level of overall physical fitness?
\nothing{img/dist/PhysicalFitness.pdf}

\noindent \textbf{Caffinated Items.} Approximately how many caffeinated foods or beverages (e.g., Coffee, Black Tea, Energy Drinks, etc.) do you consume on a regular basis?
\nothing{img/dist/CaffinatedBeverages.pdf}

\noindent \textbf{Caffeine Consumption.} Do you usually consume caffeine in the 3 hours before starting to play Beat Saber?
\nothing{img/dist/Caffeine.pdf}

\noindent \textbf{Substance Use.} How often do you play Beat Saber while under the influence of an intoxicating substance (including alcohol)?
\nothing{img/dist/SubstanceUse.pdf}

\subsection{Environment}

\noindent \textbf{Venue.} In which location do you most often play Beat Saber?
\nothing{img/dist/Venue.pdf}

\noindent \textbf{Room Size.} What are the dimensions of the play area in which you most often play Beat Saber?
\nothing{img/dist/RoomArea.pdf}

\noindent \textbf{Location.} What is the name of the country in which you most often play Beat Saber?
\nothing{img/dist/Country.pdf}

\noindent \textbf{Location.} What is the name of state or territory in which you most often play Beat Saber?
\nothing{img/dist/StateCountry.pdf}

\noindent \textbf{Location.} What is the name of the city in which you most often play Beat Saber?
\nothing{img/dist/City.pdf}

\subsection{Anthropometrics}

\nothing{img/body.png}

\noindent \textbf{Height.} What is your exact height in centimeters?
\nothing{img/dist/Height.pdf}

\noindent \textbf{Left Arm.} What is the exact length of your left arm in centimeters?
\nothing{img/dist/LeftArm.pdf}

\noindent \textbf{Right Arm.} What is the exact length of your right arm in centimeters?
\nothing{img/dist/RightArm.pdf}

\noindent \textbf{Wingspan.} What is your exact wingspan in centimeters?
\nothing{img/dist/Wingspan.pdf}

\noindent \textbf{Handedness.} Are you left or right handed?
\nothing{img/dist/Handedness.pdf}

\noindent \textbf{Weight.} What is your approximate weight in kilograms?
\nothing{img/dist/Weight.pdf}

\noindent \textbf{Interpupillary Distance.} What is your exact interpupillary distance (IPD) in millimeters?
\nothing{img/dist/IPD.pdf}

\noindent \textbf{Foot Size.} What is the exact length of your foot in centimeters?
\nothing{img/dist/FootSize.pdf}

\noindent \textbf{Hand Length.} What is the exact length of your hand in centimeters?
\nothing{img/dist/HandLength.pdf}

\noindent \textbf{Reaction Time.} What is your average reaction time in milliseconds?
\nothing{img/dist/ReactionTime.pdf}

\subsection{Clothing}

\noindent \textbf{Lower Body.} What clothing, if any, do you typically wear on your lower body when playing Beat Saber?
\nothing{img/dist/LowerBody.pdf}

\noindent \textbf{Upper Body.} What clothing, if any, do you typically wear on your upper body when playing Beat Saber?
\nothing{img/dist/UpperBody.pdf}

\noindent \textbf{Footwear.} What footwear, if any, do you typically wear when playing Beat Saber?
\nothing{img/dist/Footwear.pdf}

\eject

%% file: 910-Dist.tex
\noindent \textbf{Age} \dotfill \textbf{585} \\
\true{18-20 \dotfill ~242 (41.4\%)} \\
{21-24 \dotfill ~147 (25.1\%)} \\
{25-29 \dotfill ~80 (13.7\%)} \\
\false{30-39 \dotfill ~71 (12.1\%)} \\
\false{40-49 \dotfill ~28 (4.8\%)} \\
\false{$\geq$ 50 \dotfill ~17 (2.9\%)} \\

\noindent \textbf{AnyAthletics} \dotfill \textbf{1006} \\
\true{No \dotfill ~548 (54.5\%)} \\
\false{Yes \dotfill ~458 (45.5\%)} \\

\noindent \textbf{AnyMentalDisabilities} \dotfill \textbf{1006} \\
\true{No \dotfill ~783 (77.8\%)} \\
\false{Yes \dotfill ~223 (22.2\%)} \\

\noindent \textbf{AnyMusic} \dotfill \textbf{1006} \\
\true{No \dotfill ~558 (55.5\%)} \\
\false{Yes \dotfill ~448 (44.5\%)} \\

\noindent \textbf{AnyPhysicalDisabilities} \dotfill \textbf{1006} \\
\true{No \dotfill ~850 (84.5\%)} \\
\false{Yes \dotfill ~156 (15.5\%)} \\

\noindent \textbf{AnyRhythmGames} \dotfill \textbf{1006} \\
\true{Yes \dotfill ~660 (65.6\%)} \\
\false{No \dotfill ~346 (34.4\%)} \\

\noindent \textbf{AnyVRRhythmGames} \dotfill \textbf{1006} \\
\true{Yes \dotfill ~27 (2.7\%)} \\
\false{No \dotfill ~979 (97.3\%)} \\

\noindent \textbf{Athletics} \dotfill \textbf{858} \\
{Swimming \dotfill ~114 (13.3\%)} \\
{Soccer \dotfill ~101 (11.8\%)} \\
{Basketball \dotfill ~83 (9.7\%)} \\
\true{Tennis \dotfill ~49 (5.7\%)} \\
{Baseball \dotfill ~47 (5.5\%)} \\
\false{Track \dotfill ~36 (4.2\%)} \\
{Football \dotfill ~34 (4.0\%)} \\
{Cross Country \dotfill ~23 (2.7\%)} \\
\true{Badminton \dotfill ~20 (2.3\%)} \\
{Golf \dotfill ~19 (2.2\%)} \\
{Volleyball \dotfill ~15 (1.7\%)} \\
{Hockey \dotfill ~15 (1.7\%)} \\
{Karate \dotfill ~14 (1.6\%)} \\
{Judo \dotfill ~14 (1.6\%)} \\
{Handball \dotfill ~10 (1.2\%)} \\
{Table Tennis \dotfill ~10 (1.2\%)} \\
{Rugby \dotfill ~9 (1.0\%)} \\
{Gymnastics \dotfill ~8 (0.9\%)} \\
{Ice Hockey \dotfill ~8 (0.9\%)} \\
{Other \dotfill ~229 (25.6\%)} \\

\noindent \textbf{CaffinatedBeverages} \dotfill \textbf{974} \\
\true{None (or Rarely) \dotfill ~393 (40.3\%)} \\
{1-2 Items Weekly \dotfill ~211 (21.7\%)} \\
{1-2 Items Daily \dotfill ~278 (28.5\%)} \\
\false{3-4 Items Daily \dotfill ~61 (6.3\%)} \\
\false{5+ Items Daily \dotfill ~31 (3.2\%)} \\

\noindent \textbf{ColorBlindness} \dotfill \textbf{971} \\
\true{No \dotfill ~888 (91.5\%)} \\
\false{Yes \dotfill ~52 (5.4\%)} \\
{Maybe \dotfill ~31 (3.2\%)}

\eject

\noindent \textbf{Controller} \dotfill \textbf{1006} \\
\true{Oculus Quest \dotfill ~168 (16.7\%)} \\
\false{Valve Index \dotfill ~164 (16.3\%)} \\
{Other \dotfill ~674 (67.0\%)} \\

\noindent \textbf{Country} \dotfill \textbf{926} \\
\true{United States \dotfill ~376 (40.6\%)} \\
{Canada \dotfill ~55 (5.9\%)} \\
{United Kingdom \dotfill ~48 (5.2\%)} \\
{Australia \dotfill ~36 (3.9\%)} \\
{Germany \dotfill ~34 (3.7\%)} \\
{France \dotfill ~33 (3.6\%)} \\
{England \dotfill ~30 (3.2\%)} \\
\false{Japan \dotfill ~22 (2.4\%)} \\
{Netherlands \dotfill ~16 (1.7\%)} \\
{Finland \dotfill ~15 (1.6\%)} \\
{Poland \dotfill ~15 (1.6\%)} \\
{New Zealand \dotfill ~12 (1.3\%)} \\
{Denmark \dotfill ~11 (1.2\%)} \\
{Austria \dotfill ~10 (1.1\%)} \\
{China \dotfill ~9 (1.0\%)} \\
{Other \dotfill ~204 (22.0\%)} \\

\noindent \textbf{Dance} \dotfill \textbf{966} \\
\true{No \dotfill ~808 (83.6\%)} \\
\false{Yes, recreationally \dotfill ~134 (13.9\%)} \\
\false{Yes, professionally
or competitively \dotfill ~24 (2.5\%)} \\

\noindent \textbf{EducationalStatus} \dotfill \textbf{955} \\
\true{Less than
high school \dotfill ~345 (36.1\%)} \\
\true{High school
graduate \dotfill ~246 (25.8\%)} \\
{Some college \dotfill ~169 (17.7\%)} \\
\false{4 year degree \dotfill ~108 (11.3\%)} \\
\false{Professional
degree \dotfill ~47 (4.9\%)} \\
\false{2 year degree \dotfill ~34 (3.6\%)} \\
\false{Doctorate \dotfill ~6 (0.6\%)} \\

\noindent \textbf{EmploymentStatus} \dotfill \textbf{970} \\
\true{Student \dotfill ~521 (53.7\%)} \\
\false{Employed full time \dotfill ~229 (23.6\%)} \\
{Employed part time \dotfill ~96 (9.9\%)} \\
{Unemployed
looking for work \dotfill ~77 (7.9\%)} \\
{Unemployed not
looking for work \dotfill ~37 (3.8\%)} \\
{Disabled \dotfill ~8 (0.8\%)} \\
{Other \dotfill ~2 (0.2\%)} \\

\noindent \textbf{Ethnicity} \dotfill \textbf{976} \\
\true{White \dotfill ~760 (77.9\%)} \\
\false{Asian \dotfill ~109 (11.2\%)} \\
{Black or
African American \dotfill ~26 (2.7\%)} \\
{American Indian or
Alaska Native \dotfill ~8 (0.8\%)} \\
{Native Hawaiian
or Pacific Islander \dotfill ~3 (0.3\%)} \\
{Other \dotfill ~70 (7.2\%)} \\

\noindent \textbf{FootSize} \dotfill \textbf{811} \\
\true{$<$ 24.0 cm \dotfill ~63 (7.8\%)} \\
{24.0-24.9 cm \dotfill ~62 (7.6\%)} \\
{25.0-25.9 cm \dotfill ~140 (17.3\%)} \\
{26.0-26.9 cm \dotfill ~183 (22.6\%)} \\
{27.0-27.9 cm \dotfill ~178 (21.9\%)} \\
{28.0-28.9 cm \dotfill ~98 (12.1\%)} \\
{29.0-29.9 cm \dotfill ~39 (4.8\%)} \\
\false{$\geq$ 30.0 cm \dotfill ~48 (5.9\%)}

\eject

\noindent \textbf{Footwear} \dotfill \textbf{883} \\
\true{Typically
Barefoot \dotfill ~350 (39.6\%)} \\
{Typically
Wear Socks \dotfill ~347 (39.3\%)} \\
{Inconsistent/Varies \dotfill ~105 (11.9\%)} \\
\false{Typically
Wear Shoes \dotfill ~81 (9.2\%)} \\

\noindent \textbf{HadCOVID} \dotfill \textbf{1006} \\
\true{No \dotfill ~655 (65.1\%)} \\
\false{Yes \dotfill ~351 (34.9\%)} \\

\noindent \textbf{HandLength} \dotfill \textbf{760} \\
\true{$<$ 16.0 cm \dotfill ~41 (5.4\%)} \\
{16.0-16.9 cm \dotfill ~57 (7.5\%)} \\
{17.0-17.9 cm \dotfill ~132 (17.4\%)} \\
\false{18.0-18.9 cm \dotfill ~151 (19.9\%)} \\
\false{19.0-19.9 cm \dotfill ~185 (24.3\%)} \\
\false{20.0-20.9 cm \dotfill ~99 (13.0\%)} \\
\false{21.0-21.9 cm \dotfill ~32 (4.2\%)} \\
\false{$\geq$ 22.0 cm \dotfill ~63 (8.3\%)} \\

\noindent \textbf{Handedness} \dotfill \textbf{900} \\
\true{Right Handed \dotfill ~737 (81.9\%)} \\
\false{Left Handed \dotfill ~103 (11.4\%)} \\
{Ambidextrous \dotfill ~60 (6.7\%)} \\

\noindent \textbf{Headset} \dotfill \textbf{1006} \\
\true{Oculus Quest 2 \dotfill ~499 (49.6\%)} \\
\false{Valve Index \dotfill ~150 (14.9\%)} \\
{Other \dotfill ~357 (35.5\%)} \\

\noindent \textbf{Height} \dotfill \textbf{838} \\
\true{$<$ 1.70 m \dotfill ~191 (22.8\%)} \\
{1.70-1.79 m \dotfill ~321 (38.3\%)} \\
{1.80-1.89 m \dotfill ~288 (34.4\%)} \\
\false{$\geq$ 1.90 m \dotfill ~38 (4.5\%)} \\

\noindent \textbf{IPD} \dotfill \textbf{737} \\
\true{$<$ 58.0 mm \dotfill ~28 (3.8\%)} \\
{58.0-62.9 mm \dotfill ~153 (20.8\%)} \\
{63.0-67.9 mm \dotfill ~373 (50.6\%)} \\
{68.0-71.9 mm \dotfill ~143 (19.4\%)} \\
\false{$\geq$ 72.0 mm \dotfill ~40 (5.4\%)} \\

\noindent \textbf{Income} \dotfill \textbf{908} \\
\true{Less than \$10,000 \dotfill ~583 (64.2\%)} \\
\false{\$10,000 to \$19,999 \dotfill ~94 (10.4\%)} \\
\false{\$20,000 to \$29,999 \dotfill ~39 (4.3\%)} \\
\false{\$30,000 to \$39,999 \dotfill ~45 (5.0\%)} \\
\false{\$40,000 to \$49,999 \dotfill ~39 (4.3\%)} \\
\false{\$50,000 to \$59,999 \dotfill ~25 (2.8\%)} \\
\false{\$60,000 to \$69,999 \dotfill ~14 (1.5\%)} \\
\false{\$70,000 to \$79,999 \dotfill ~14 (1.5\%)} \\
\false{\$80,000 to \$89,999 \dotfill ~16 (1.8\%)} \\
\false{\$90,000 to \$99,999 \dotfill ~5 (0.6\%)} \\
\false{\$100,000 to \$149,999 \dotfill ~20 (2.2\%)} \\
\false{More than \$150,000 \dotfill ~14 (1.5\%)} \\

\noindent \textbf{Languages} \dotfill \textbf{1406} \\
\true{English \dotfill ~845 (60.1\%)} \\
{German \dotfill ~71 (5.0\%)} \\
{French \dotfill ~58 (4.1\%)} \\
\false{Spanish \dotfill ~49 (3.5\%)} \\
{Japanese \dotfill ~37 (2.6\%)} \\
{Dutch \dotfill ~33 (2.3\%)} \\
{Polish \dotfill ~22 (1.6\%)} \\
{Other \dotfill ~270 (19.2\%)}

\eject

\noindent \textbf{LeftArm} \dotfill \textbf{692} \\
\true{$<$ 0.60 m \dotfill ~62 (9.0\%)} \\
{0.60-0.69 m \dotfill ~256 (37.0\%)} \\
{0.70-0.79 m \dotfill ~306 (44.2\%)} \\
\false{$\geq$ 0.80 m \dotfill ~68 (9.8\%)} \\

\noindent \textbf{Lenses} \dotfill \textbf{972} \\
\true{Never \dotfill ~721 (74.2\%)} \\
\false{Often \dotfill ~231 (23.8\%)} \\
{Sometimes \dotfill ~20 (2.1\%)} \\

\noindent \textbf{LowerBody} \dotfill \textbf{880} \\
\true{Knee-Height Garment \dotfill ~350 (39.8\%)} \\
{Ankle-Height Garment \dotfill ~248 (28.2\%)} \\
{Inconsistent/Varies \dotfill ~194 (22.0\%)} \\
\false{Undergarment Only \dotfill ~88 (10.0\%)} \\

\noindent \textbf{MaritalStatus} \dotfill \textbf{959} \\
\true{Never married \dotfill ~896 (93.4\%)} \\
\false{Married \dotfill ~39 (4.1\%)} \\
\false{Divorced \dotfill ~16 (1.7\%)} \\
\false{Separated \dotfill ~6 (0.6\%)} \\
\false{Widowed \dotfill ~2 (0.2\%)} \\

\noindent \textbf{Music} \dotfill \textbf{714} \\
\true{Piano \dotfill ~195 (27.3\%)} \\
\true{Guitar \dotfill ~107 (15.0\%)} \\
{Drums \dotfill ~63 (8.8\%)} \\
\false{Trumpet \dotfill ~58 (8.1\%)} \\
{Violin \dotfill ~41 (5.7\%)} \\
\false{Saxophone \dotfill ~26 (3.6\%)} \\
\false{Trombone \dotfill ~23 (3.2\%)} \\
\false{Clarinet \dotfill ~21 (2.9\%)} \\
\false{Flute \dotfill ~18 (2.5\%)} \\
{Cello \dotfill ~16 (2.2\%)} \\
{Bass \dotfill ~14 (2.0\%)} \\
{Recorder \dotfill ~11 (1.5\%)} \\
{Ukulele \dotfill ~8 (1.1\%)} \\
{Viola \dotfill ~7 (1.0\%)} \\
{Percussion \dotfill ~6 (0.8\%)} \\
{French Horn \dotfill ~5 (0.7\%)} \\
{Tuba \dotfill ~4 (0.6\%)} \\
{Other \dotfill ~91 (12.7\%)} \\

\noindent \textbf{PhysicalFitness} \dotfill \textbf{972} \\
\true{Far below average \dotfill ~51 (5.2\%)} \\
{Below average \dotfill ~252 (25.9\%)} \\
{Average \dotfill ~387 (39.8\%)} \\
{Above average \dotfill ~242 (24.9\%)} \\
\false{Far above average \dotfill ~40 (4.1\%)} \\

\noindent \textbf{PoliticalOrientation} \dotfill \textbf{934} \\
{Independent or Neither \dotfill ~454 (48.6\%)} \\
\true{Liberal or Left Wing \dotfill ~334 (35.8\%)} \\
\false{Conservative or Right Wing \dotfill ~72 (7.7\%)} \\
{Other \dotfill ~74 (7.9\%)} \\

\noindent \textbf{Preparation} \dotfill \textbf{1006} \\
\true{None \dotfill ~247 (24.6\%)} \\
\false{Warmup Only \dotfill ~244 (24.3\%)} \\
{Warmup and Stretches \dotfill ~192 (19.1\%)} \\
{Other \dotfill ~323 (32.1\%)}

\eject

\noindent \textbf{ReactionTime} \dotfill \textbf{852} \\
\true{$<$ 125 ms \dotfill ~8 (0.9\%)} \\
\true{125-174 ms \dotfill ~59 (6.9\%)} \\
{175-199 ms \dotfill ~166 (19.5\%)} \\
{200-224 ms \dotfill ~233 (27.3\%)} \\
{225-274 ms \dotfill ~230 (27.0\%)} \\
{275-324 ms \dotfill ~103 (12.1\%)} \\
\false{$\geq$ 325 ms \dotfill ~53 (6.2\%)} \\

\noindent \textbf{RhythmGames} \dotfill \textbf{1850} \\
{Osu \dotfill ~383 (20.7\%)} \\
{Guitar Hero \dotfill ~107 (5.8\%)} \\
{A Dance Of  Fire And Ice \dotfill ~121 (6.5\%)} \\
{Geometry Dash \dotfill ~100 (5.4\%)} \\
\true{Quaver \dotfill ~82 (4.4\%)} \\
{Muse Dash \dotfill ~65 (3.5\%)} \\
\false{Dance Dance Revolution \dotfill ~42 (2.3\%)} \\
{Synth Riders \dotfill ~38 (2.1\%)} \\
{Arcaea \dotfill ~36 (1.9\%)} \\
{Pistol Whip \dotfill ~27 (1.5\%)} \\
{Rock Band \dotfill ~25 (1.4\%)} \\
{Sound Voltex \dotfill ~22 (1.2\%)} \\
{Audica \dotfill ~22 (1.2\%)} \\
{Rhythm Doctor \dotfill ~18 (1.0\%)} \\
{Osumania \dotfill ~18 (1.0\%)} \\
{Spin Rhythm Xd \dotfill ~18 (1.0\%)} \\
{Wacca \dotfill ~17 (0.9\%)} \\
{Phigros \dotfill ~17 (0.9\%)} \\
{Just Dance \dotfill ~14 (0.8\%)} \\
{Stepmania \dotfill ~14 (0.8\%)} \\
{Clone Hero \dotfill ~13 (0.7\%)} \\
{Trombone Champ \dotfill ~13 (0.7\%)} \\
{Boombox \dotfill ~13 (0.7\%)} \\
{Other \dotfill ~625 (33.8\%)} \\

\noindent \textbf{RightArm} \dotfill \textbf{692} \\
\true{$<$ 0.60 m \dotfill ~57 (8.2\%)} \\
{0.60-0.69 m \dotfill ~256 (37.0\%)} \\
{0.70-0.79 m \dotfill ~305 (44.1\%)} \\
\false{$\geq$ 0.80 m \dotfill ~74 (10.7\%)} \\

\noindent \textbf{RoomArea} \dotfill \textbf{589} \\
\true{$<$ 2.0 m² \dotfill ~82 (13.9\%)} \\
{2.0-3.9 m² \dotfill ~187 (31.7\%)} \\
{4.0-5.9 m² \dotfill ~161 (27.3\%)} \\
{6.0-7.9 m² \dotfill ~99 (16.8\%)} \\
\false{$\geq$ 8.0 m² \dotfill ~60 (10.2\%)} \\

\noindent \textbf{Sex} \dotfill \textbf{979} \\
\true{Male \dotfill ~806 (82.3\%)} \\
\false{Female \dotfill ~91 (9.3\%)} \\
{Other \dotfill ~82 (8.4\%)} \\

\noindent \textbf{StandaloneGrip} \dotfill \textbf{475} \\
\true{Default Grip \dotfill ~229 (48.2\%)} \\
\false{Claw Grip \dotfill ~154 (32.4\%)} \\
{Standard M-Grip \dotfill ~13 (2.7\%)} \\
{Yoshi M-Grip \dotfill ~10 (2.1\%)} \\
{Standard C-Grip \dotfill ~7 (1.5\%)} \\
{Crab Grip \dotfill ~6 (1.3\%)} \\
{Not Applicable
/ Varies \dotfill ~4 (0.8\%)} \\
{Other \dotfill ~52 (10.9\%)} \\

\noindent \textbf{SteamComputerFormFactor} \dotfill \textbf{568} \\
\true{Desktop \dotfill ~513 (90.3\%)} \\
\false{Laptop \dotfill ~55 (9.7\%)}

\eject

\noindent \textbf{SteamLighthouses} \dotfill \textbf{194} \\
\true{1 \dotfill ~13 (6.7\%)} \\
{2 \dotfill ~133 (68.6\%)} \\
\false{3 \dotfill ~35 (18.0\%)} \\
\false{4 \dotfill ~13 (6.7\%)} \\

\noindent \textbf{SteamOperatingSystemVersion} \dotfill \textbf{574} \\
\true{Windows 10 (64 bit) \dotfill ~385 (67.1\%)} \\
\false{Windows 11 (64 bit) \dotfill ~177 (30.8\%)} \\
{Other \dotfill ~12 (2.1\%)} \\

\noindent \textbf{SteamProcessorCPUVendor} \dotfill \textbf{572} \\
\true{AMD \dotfill ~333 (58.2\%)} \\
\false{Intel \dotfill ~239 (41.8\%)} \\

\noindent \textbf{SteamProcessorLogicalCores} \dotfill \textbf{569} \\
\true{4 \dotfill ~8 (1.4\%)} \\
\true{6 \dotfill ~22 (3.9\%)} \\
\true{8 \dotfill ~48 (8.4\%)} \\
{12 \dotfill ~199 (35.0\%)} \\
\false{16 \dotfill ~199 (35.0\%)} \\
\false{20 \dotfill ~30 (5.3\%)} \\
\false{24 \dotfill ~45 (7.9\%)} \\
\false{32 \dotfill ~18 (3.2\%)} \\

\noindent \textbf{SubstanceUse} \dotfill \textbf{970} \\
\true{Never \dotfill ~778 (80.2\%)} \\
{Rarely \dotfill ~130 (13.4\%)} \\
{Somewhat Often \dotfill ~24 (2.5\%)} \\
\false{Often \dotfill ~38 (3.9\%)} \\

\noindent \textbf{TotalPlayTime} \dotfill \textbf{1006} \\
\true{$<$ 100 Hours \dotfill ~200 (19.9\%)} \\
{100-499 Hours \dotfill ~358 (35.6\%)} \\
{500-999 Hours \dotfill ~208 (20.7\%)} \\
\false{1000-1999 Hours \dotfill ~182 (18.1\%)} \\
\false{$\geq$ 2000 Hours \dotfill ~58 (5.8\%)} \\

\noindent \textbf{UpperBody} \dotfill \textbf{764} \\
\true{Short Sleeve Garment \dotfill ~526 (68.8\%)} \\
{Inconsistent/Varies \dotfill ~133 (17.4\%)} \\
{Sleeveless Garment \dotfill ~52 (6.8\%)} \\
{Undergarment Only \dotfill ~23 (3.0\%)} \\
\false{Long Sleeve Garment \dotfill ~20 (2.6\%)} \\
\false{Multiple Layers \dotfill ~10 (1.3\%)} \\

\noindent \textbf{Weight} \dotfill \textbf{834} \\
\true{$<$ 40.0 kg \dotfill ~5 (0.6\%)} \\
\true{40.0-49.9 kg \dotfill ~51 (6.1\%)} \\
{50.0-59.9 kg \dotfill ~154 (18.5\%)} \\
{60.0-69.9 kg \dotfill ~189 (22.7\%)} \\
{70.0-79.9 kg \dotfill ~193 (23.1\%)} \\
{80.0-89.9 kg \dotfill ~103 (12.4\%)} \\
{90.0-99.9 kg \dotfill ~70 (8.4\%)} \\
\false{$\geq$ 100.0 kg \dotfill ~69 (8.3\%)} \\

\noindent \textbf{Wingspan} \dotfill \textbf{710} \\
\true{$<$ 1.60 m \dotfill ~133 (18.7\%)} \\
{1.60-1.69 m \dotfill ~149 (21.0\%)} \\
{1.70-1.79 m \dotfill ~205 (28.9\%)} \\
{1.80-1.89 m \dotfill ~167 (23.5\%)} \\
\false{$\geq$ 1.90 m \dotfill ~56 (7.9\%)}

\eject